\documentclass[showpacs,floatfix,superscriptaddress,showpacs,twocolumn,amssymb,amsfonts,prb,aps]{revtex4}
\usepackage{longtable,graphicx,epsfig,dcolumn}

\begin{document}
\bibliographystyle{revtex}
\title{Solvent Mediated Assembly of Nanoparticles Confined in Mesoporous Alumina}

\author{Kyle~J.~Alvine}
\email{alvine@fas.harvard.edu} \affiliation{Division of
Engineering and Applied Sciences, Harvard University, Cambridge MA
02138, USA}

\author{Diego~Pontoni}
\affiliation{Division of Engineering and Applied Sciences, Harvard
University, Cambridge MA 02138, USA}

\author{Oleg~G.~Shpyrko}
\affiliation{Division of Engineering and Applied Sciences, Harvard
University, Cambridge MA 02138, USA}\affiliation{The Center for
Nanoscale Materials, Argonne National Laboratory, Argonne, IL,
60439, USA}

\author{Peter~S.~Pershan}
\affiliation{Division of Engineering and Applied Sciences, Harvard
University, Cambridge MA 02138, USA}\affiliation{Department of
Physics, Harvard University, Cambridge MA 02138, USA}

\author{David~J.~Cookson}
\affiliation{Australian Synchrotron Research Program, Bldg 434,
Argonne National Laboratory, Argonne, IL, 60439, USA}

\author{Kyusoon~Shin}
\altaffiliation[Current address:  ]{School of Chemical and
Biological Engineering, Seoul National University, Seoul, South
Korea}
\author{Thomas~P.~Russell}
\affiliation{Department of Polymer Science and
Engineering,University of Massachusetts, Amherst MA 01003, USA}

\author{Markus Brunnbauer}
\author{Francesco Stellacci}
\affiliation{Department of Materials Science and Engineering,
Massachusetts Institute of Technology, Cambridge, MA 02138, USA}

\author{Oleg~Gang}
\affiliation{Center for Functional Nanomaterials, Brookhaven
National Lab, Upton NY 11973, USA}

\date{Submitted 20 December 2005}

\begin{abstract}

\def\baselinestretch{1}
\noindent  The controlled self-assembly of thiol stabilized gold
nanocrystals in a mediating solvent and confined within mesoporous
alumina was probed \emph{in-situ} with small angle x-ray
scattering.  The evolution of the self-assembly process was
controlled reversibly via regulated changes in the amount of
solvent condensed from an under-saturated vapor.  Analysis
indicated that the nanoparticles self-assembled into cylindrical
monolayers within the porous template. Nanoparticle
nearest-neighbor separation within the monolayer increased and the
ordering decreased with the controlled addition of solvent.  The
process was reversible with the removal of solvent.  Isotropic
clusters of nanoparticles were also observed to form temporarily
during desorption of the liquid solvent and disappeared upon
complete removal of liquid. Measurements of the absorption and
desorption of the solvent showed strong hysteresis upon thermal
cycling. In addition, the capillary filling transition for the
solvent in the nanoparticle-doped pores was shifted to larger
chemical potential, relative to the liquid/vapor coexistence, by a
factor of four as compared to the expected value for the same
system without nanoparticles.

\end{abstract}

\pacs{61.46.Df, 68.08.Bc, 61.10.Eq }
\maketitle
\section{INTRODUCTION}Structures of self-assembled nanoparticles
have been studied extensively in recent years for their unique
catalytic, electronic, and optical properties. The geometry of
self-assembled structures can vary from 2D sheets of
nanoparticles, to spherical monolayers or 1D nanowire
arrangements.\cite{Narayanan04, Benfield01, Sawitowski01,
Lu05,Dokoutchaev99} Reduced dimensionality systems, in particular
nanowires, have been of recent interest for their potential
electronic and optical properties.\cite{Hu99, Sawitowski01} There
has been additional interest in quasi-one dimensional structures,
such as the arrangement of nanoparticles into cylindrical
monolayers, for their similarity with biological systems such as
some virus protein coatings and cell
microtubules.\cite{Erickson73, Liu05} Understanding the process of
nanoparticle self-assembly in the different geometries can be
vital to production of functional nanoscale structures. Recently
there have been several experimental and theoretical studies that
examine the nanoparticle self-assembly on flat surfaces via the
evaporation of macroscopic droplets. \cite{Narayanan04,
Narayanan05, Rabani03, Lu05,Ohara95, Lin01,Ge00}  One dimensional
systems typically require templates, thus bulk methods, such as
droplet evaporation, are not possible. Recent studies of quasi-1D
nanoparticle structures within cylindrical
templates\cite{Benfield01, Sawitowski01, Lahav03, Mickelson03,
Kelberg03, Bronstein03, Konya03} primarily probed the end product
and did not explicitly measure the evolution of the self-assembly
process. In order to investigate the self assembly process itself
within confined geometries, it is necessary to perform
\emph{in-situ} experiments with precise control of the solvent
amount, analogous to controlled evaporation of macroscopic
droplets of nanoparticles on flat surfaces.

    Here we describe \emph{in-situ} small angle x-ray scattering (SAXS) experiments
that probed the solvent mediated self-assembly of Au-core,
colloidal nanoparticles within nanoporous alumina.  Experiments
were carried out within an environmental chamber which allowed
precise control over the amount of solvent condensed from vapor
into the porous system. Analysis determined that the nanoparticles
self-assembled into a cylindrical monolayer that evolved with the
addition and removal of liquid.  The evolution of this cylindrical
monolayer structure was completely reversible upon removal of
liquid, albeit with strong hysteresis typical in capillary
systems.  In addition to the cylindrical monolayer, isotropic
clusters temporarily formed during desorption of the liquid. These
clusters disappeared upon complete removal of the liquid from the
pores.

\section{EXPERIMENTAL} 
\begin{figure}[tbp]
  \includegraphics[width=1\columnwidth]{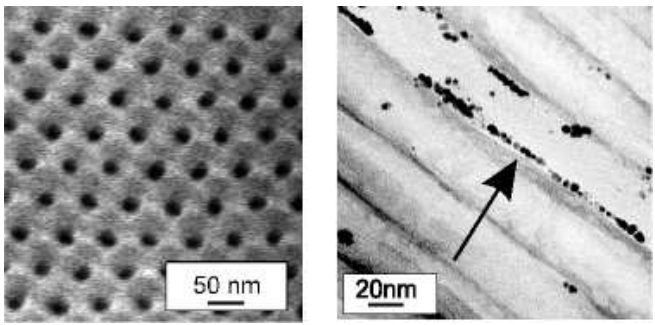}
  \caption{(LEFT) SEM image shows well ordered hexagonal packing of alumina
  nanopores prepared under similar conditions to samples used for x-ray experiments.
  (RIGHT) Bright-field TEM image of the nanoparticles (arrow) on the wall (faint diagonal lines) of alumina
  pores. The nanoparticles are Au-core with octane-thiol coating.
  Cross-sections were prepared by ultra micro-tome method
  from the sample used for x-ray experiments.} \label{fig:tem}
\end{figure}

Anodized alumina membranes were electrochemically
prepared using a two-step anodization technique described
elsewhere. \cite{Masuda95, Shin04, Xiang05} The aluminum backing
layer was then dissolved in HgCl$_2$.  Pores were opened by
floating on phosphoric acid followed by thorough rinsing in
de-ionized water. The resultant alumina membrane consisted of a
dense matrix of pores running completely through the membrane and
perpendicular to its surface. The long axes of the pores were
parallel and arranged in a 2D powder-like arrangement with local
hexagonal packing.  Nearest neighbor distances (center to center)
of 63$\pm$2~nm were determined via x-ray scattering.  This is
consistent with electron microscopy of similarly prepared samples
(see Fig.~\ref{fig:tem}). Alumina nanopores were 29$\pm$4~nm in
diameter (from TEM). The macroscopic dimensions of the nanoporous
membrane were about 1~cm~$\times$~1~cm~$\times$~90~microns.

    Nanoporous alumina membranes were further cleaned in solvents
to remove organic impurities, with 15 minute ultra-sonic baths in
each of the following (in order): HPLC grade chloroform, 99.7\%
pure acetone, HPLC grade methanol, and HPLC grade toluene.  The
membrane was then allowed to soak in HPLC grade toluene for 24
hours to remove any further impurities. After soaking, the
membrane was transferred without drying to 2~ml of a 2.26~mg/ml
solution of octane-thiol (OT) stabilized Au-core
nanocrystals\cite{Jackson04, Terrill95} in HPLC grade toluene for
two weeks at 25$^\circ$C.  Fits of SAXS data from bulk scattering
of a dilute suspension of the same solution of nanoparticles used
in doping the nanopores yielded $R_s=1.2$~nm and
$\Delta_R=0.16$~nm giving a polydispersity of $\Delta_R/R_s=13\%$.
Since the electron density of the Au core is much higher than the
organic ligands, $R_s$ was primarily a measure of the core radius.

    The nanoparticle solution was drawn into the alumina pores via
capillary forces\cite{Benfield01, Sawitowski01}. The initial
concentration (and volume) of the nanoparticle solution was chosen
such that the total number of particles in the initial solution
was about 2-3 times the number of nanoparticles needed to form a
complete monolayer of nanoparticles along the walls of the pores
(assuming hexagonal close packing), but less than that required
for complete volume filling of the pores. In addition to soaking,
the solution with the immersed membrane was placed in an
ultra-sonic bath for 15 minutes per day to facilitate movement of
the particles into the membrane.  The long time period and
frequent ultra-sonic baths were imposed to ensure maximal
integration of nanoparticles into the porous membranes.

\begin{figure}[tbp]
  \includegraphics[width=.8\columnwidth]{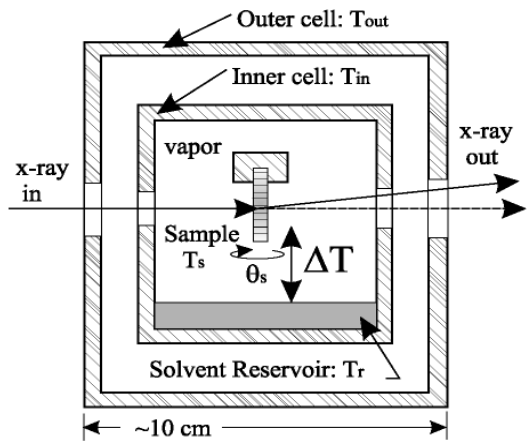}
  \caption{Schematic of the environmental chamber used for \emph{in-situ} x-ray experiments.
  The outer (inner) cell was kept at 28$^\circ$C (32$^\circ$C) during the experiment.
  The sample temperature $T_s$ was controlled independently from the inner cell and liquid
  solvent reservoir temperature, $T_r$.  The sample was mounted in a transmission geometry
  where the long axis of the pores can be rotated ($\theta_s$) in the horizontal plane.
  The $\Delta T=T_s-T_r$ was controlled to vary liquid condensation on the sample via the vapor.}
  \label{fig:cell}
\end{figure}

 After 24 hours in solution the alumina membrane color changed
from clear to dark red color, similar to the solution color.  This
was an indication that nanoparticles had been absorbed into the
membrane.  The presence of nanoparticles in the pores was
confirmed with cross-sectional TEM of the sample (see
Fig.~\ref{fig:tem}). No further color changes of the membrane were
observed. Additionally, no significant color change of the
solution was observed, indicating that some particles were left in
the solution.  The sample was then removed from the solution and
allowed to dry in air for five minutes before being loaded into an
environmental chamber\cite{Tidswell91, Heilmann01} (see
Fig.~\ref{fig:cell}) for the SAXS experiment.

    Prior to sample mounting, all components of the inner cell of the environmental
chamber were cleaned with solvents: 15 minute ultra-sonic baths of
each of HPLC grade chloroform, 99.7\% pure acetone, and HPLC grade
methanol.  During the x-ray experiment the inner cell was
hermetically sealed and the temperatures of the inner and outer
cells were kept at $T_{in}=32.0\pm0.5^\circ$~C and
$T_{out}=28.0\pm0.5^\circ$~C, respectively.  The actual
temperature stability during each measurement was better than $\pm
50$~mK, though the temperature of the cells did vary $\pm 0.5$~K
between different measurements due to the differing  heat loads on
the sample as described below (see the description of $\Delta T$).
The sample mounting block was thermally isolated to a high degree
from the walls of the inner cell with independent thermal control
via a heater on the mounting block. All temperatures were
continuously monitored with YSI \#45008 thermistors.
    A liquid solvent reservoir was created by injecting 10~ml of HPLC
grade liquid toluene into the inner cell. Solvent condensation in
the sample was precisely controlled via the (positive) offset
between the sample temperature, $T_s$, and the liquid solvent
reservoir temperature, $T_r$: $\Delta T=T_s-T_r$\cite{Tidswell91,
Heilmann01}.  The chemical potential offset (from the liquid/vapor
coexistence) is given by: $\Delta\mu\approx H_{vap}\Delta T/T_r$.
Here, $H_{vap}=38.06$~kJ/mol is the heat of vaporization of
toluene\cite{Majer85} and $T_r$ was kept constant throughout the
experiment (as described above) via temperature control of the
inner cell.  Liquid was injected into the cell at a large $\Delta
T\approx 30$~K to avoid rapid condensation in the pores.  Since
$\Delta T$ could be varied from about 50~mK up to 30~K, the
corresponding $\Delta\mu$ could be varied about four orders of
magnitude.  As $\Delta T$ and thus $\Delta\mu $ decreased the
amount of condensed solvent increased. Likewise, solvent was
removed by increasing $\Delta T$.  In this experiment, the pores
became saturated with liquid for all $\Delta T < 4$~K, thus the
experimental range of $\Delta T$ probed was only 0.5~K$<\Delta
T<30$~K.  X-rays were allowed to pass through the chamber via
0.02~mm Kapton\copyright~(Dupont) windows on the outer chamber and
0.005 inch thick beryllium windows on the inner chamber.

\section{SMALL ANGLE X-RAY SCATTERING} 
\begin{figure}[tbp]
  \includegraphics[width=1.0\columnwidth]{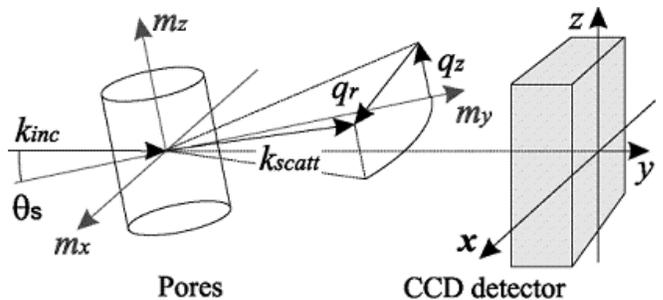}
  \caption{Geometry for SAXS measurements:  X-rays were incident at $\theta_s=10^\circ$
  from normal to the nanopore long axis.  Scattered intensity
  was measured with a fixed position CCD detector down-stream.
  The $\vec{m}$ and $\vec{q}$ describe the coordinate space and wavevector transfer
  of the nanopore, respectively.  Note that the geometry shown here has
  been rotated about the incident beam by $90^\circ$ for clarity.} \label{fig:geom}
\end{figure}

\emph{In-situ}
x-ray measurements were carried out at the SAXS facility at
ChemMatCARS beamline at the Advanced Photon Source (Argonne
National Lab, Argonne, IL). The incident x-ray energy of 11.55~keV
was well below the L3 absorption edge of Au (11.92~keV) to avoid
fluorescence. The sample chamber was mounted on a goniometer in a
transmission geometry with two degrees of rotational freedom --
sample rotation and tilt -- plus the standard three translations.
The sample rotation was done internally and allowed rotation
angles, $\theta_s$, of the sample with respect to the incident
beam in excess of $\pm 90^\circ$ (see Fig.~\ref{fig:cell}). A
fixed position geometry CCD detector measured the SAXS intensity
at a camera length, $L$, of 1880~mm down-stream from the sample.
Powder diffraction rings were seen at small angles due to the 2D
hexagonal packing of the nanopores. Sample alignment was performed
with the membrane nearly perpendicular to the incident beam and
thus the pores were parallel to the incident beam.  The sample
rotation and tilt were adjusted to maximize the symmetry of the
diffraction rings in both the vertical and horizontal directions.
After alignment, the membrane was exactly perpendicular to the
incident beam and the long axes of the pores were parallel to the
incident beam.  This defined $\theta_s=90^\circ$ between the
incident beam and the short axis of the pores.  The membrane was
then rotated to $\theta_s=10\pm1^\circ$ from the incident beam,
i.e. the pore long axis was then at 80$^\circ$ to the incident
x-rays (see Fig.~\ref{fig:geom}). This geometry allowed a bulk
measurement of the nanoparticle/nanopore system while maximizing
the wavevector transfer along the nanopore long axis, $q_z$, that
was probed.

    It is important to note that the geometry
shown in Fig.~\ref{fig:geom} is rotated by $90^\circ$ about the
incident beam for ease of viewing.  In the figure, the long axis
of the pores is depicted as approximately perpendicular to the
incident beam and vertical with rotations $\theta_s$ in the
vertical plane.  In the actual experiment, the only difference was
that the long axis of the pore was horizontal (though still
perpendicular to the incident beam) and rotations $\theta_s$ were
done in the horizontal plane.
\section{Volume of adsorbed solvent} 
\begin{figure}[tbp]
\includegraphics[width=1.0\columnwidth]{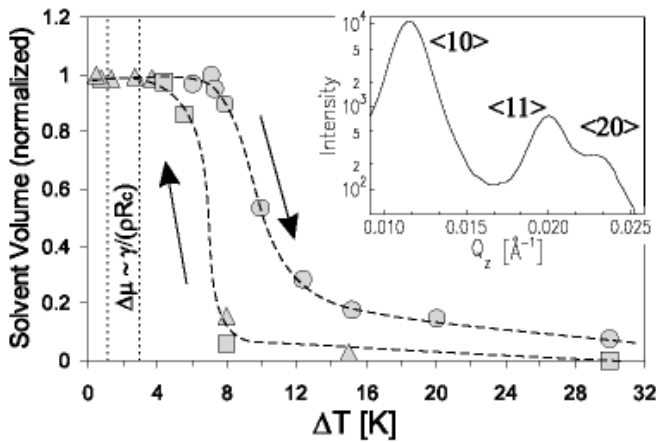}
\caption{(Main) Volume adsorption($\bigtriangleup,\Box$) /
desorption($\bigcirc$) curves as a function of $\Delta T$.  Dashed
lines are added as a guide for the eye. Note the strong hysteresis
upon cycling.  Vertical dotted lines indicate location of
capillary transition for empty (no nanoparticles) pores predicted
by the Kelvin equation, $\Delta\mu\sim\gamma/\rho R_c$. Data is
normalized to the liquid saturation volume of the pores. (Inset)
Plot showing the powder diffraction peaks from the 2D nanopore
packing.  Volume information was extracted from the first peak
(the $\langle 10\rangle$).}\label{fig:lowq}
\end{figure}

A measure of absorbed volume of solvent in the pores
was calculated from the small angle powder diffraction peaks
mentioned above. As liquid was condensed into the pores, the
electron density contrast in the pores decreased relative to that
of the dry pores, thus reducing the scattering intensity of the
diffraction peaks, $I_{peak}(\Delta T)$, relative to the dry
peaks, $I_{peak}(dry)$. Neglecting absorption corrections, the
added solvent volume, $V_{liq}$, was then approximately determined
from the lowest order peak (small angle approximation)
via\cite{volume}:
\begin{eqnarray}
V_{liq}(\Delta
T)&\propto&\sqrt{I_{peak}(dry)}-\sqrt{I_{peak}(\Delta T)}
\end{eqnarray}
Plots of the added volume derived from the $\langle 10\rangle$
diffraction peak are shown in Fig.~\ref{fig:lowq}.
The data represent three different thermal cycles:  two cooling
cycles (large $\Delta T$ to small $\Delta T$) and one heating
cycle.  The data have been normalized to the saturated (pores
completely filled with liquid) volume for each curve. There are
two main features in this figure: the first is the presence of a
sharp transition (from low relative volume to saturation and
reverse) for both adsorption and desorption curves and second,
there is a marked hysteresis which is reproducible (at least for
the cooling cycle).  For both sets of curves, the
nanopore/nanoparticle system is saturated with toluene for $\Delta
T \leq4$~K and has negligible amounts of toluene for $\Delta T
\geq30$~K.  These two regions will be referred to throughout the
rest of the paper as the ``saturated'' and ``dry'' regions,
respectively.

    The simplest theory to describe the observed capillary
transition is the Kelvin equation:
\begin{eqnarray}
\Delta \mu_{adsorption}&\approx&\frac{\gamma}{n_lR_c}\\
\Delta \mu_{desorption}&\approx&\frac{2\gamma}{n_lR_c}
\end{eqnarray}
where $R_c$ is the cylindrical nanopore radius, $\gamma=28.4$~mN/m
is the surface tension of bulk toluene, and
$n_l=\rho/m_{wt}=9\times 10^{-3}$~moles/cm$^3$ (mass density
divided by molecular weight) is the molar density of the toluene.
The chemical potential at the capillary transition for the
adsorption and desorption differ by a factor of two due to
different physical mechanisms for adsorption and desorption. Pores
are expected to fill by coaxial film growth (curvature of $1/R_c$)
and empty from the ends of the pores via a spherical meniscus
(curvature of $2/R_c$) that travels the length of the
pore.\cite{Cohan38} This argument provides a motivation for the
presence of the hysteresis, even though the system studied is
clearly more complicated due to the presence of the nanoparticles.
According to the Kelvin equation, the observed capillary filling
transition in Fig.~\ref{fig:lowq} would indicate an effective
radius for the pores of about 4~nm, almost a factor of four
smaller than the actual pore radius of 15~nm.  For similar
nanoporous alumina samples with no nanoparticles, it was found
that the Kelvin equation provides a good prediction of the
capillary transition\cite{AlvineUP}. Significantly higher values
of $\Delta T$ for the capillary transition indicated that there
was a real reduction in the effective radius of the pores due to
the presence of the nanoparticles.  Assuming a monolayer thickness
due to the nanoparticles of 2.4~nm for the Au core plus
1.2~nm$\times$2 for the organic OT shell\cite{Bain89}$\approx
5$~nm, this would bring the effective radius down to about 10~nm.
Additionally, roughness from the nanoparticle monolayer surface
contributes to the transition shift by increasing the amount of
adsorbed liquid at large $\Delta T$\cite{Robbins91}.

\section{Nanoparticle-nanoparticle Scattering}
\subsection{Elliptical Transforms}
\begin{figure}[tbp]
\includegraphics[width=1\columnwidth]{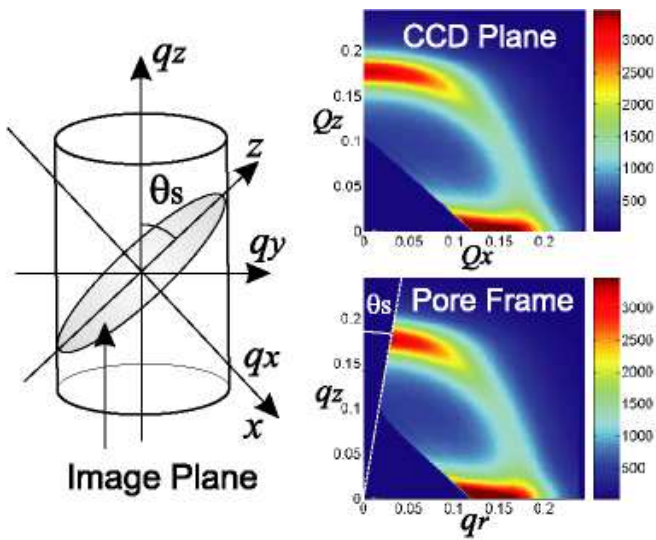}
\caption{(Color)(Left) Schematic of elliptical transforms of
scattering intensity data.  Lines of constant $q_r$ are ellipses
in CCD coordinates where the scattering (CCD)
 plane intersects the cylindrical surfaces of constant $q_r$.
 (Right Top) CCD image of scattering intensity at $\Delta T=30$~K.
 The direct beam is in the lower left corner.  The
 triangular region here is due to an attenuator used to block the
 intense pore-pore scattering at low $q$.
 (Right Bottom) The same data set transformed to the $\vec{q}$ coordinates
 relative to the nanopore axis. }\label{fig:transform}
\end{figure}

\begin{figure}[tbp]
  \includegraphics[width=1.0\columnwidth]{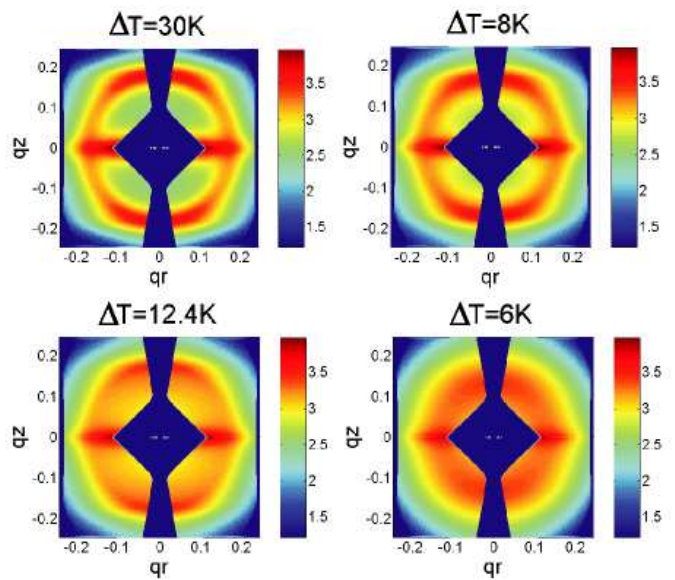}
  \caption{(Color)Scattering intensity distributions at four representative $\Delta T$.
Data (intensity log scale) were background subtracted, then tiled
into all quadrants to simulate a full ring.  (Top Left) Data from
$\Delta T=30$~K (dry): the scattering intensity is an asymmetric
ring structure from particle-particle scattering. (Top Right) Data
from $\Delta T=8$~K (adsorption curve): note the ring is broader
with a smaller radius. (Bottom Right)  Data from $\Delta T=6$~K
(saturated with liquid): here the ring is even broader with an
even smaller radius. (Bottom Left) Data from $\Delta T=12.4$~K
(desorption curve): note the presence of an isotropic scattering
ring just inside the asymmetric ring.} \label{fig:hiq}
\end{figure}

In addition to the low angle powder diffraction from the 2D
hexagonal pore packing, a ring-like structure was observed at
larger angles ($q\approx0.18$~$\dot{A}^{-1}$, see
Fig.~\ref{fig:transform}), which corresponds to nanoparticle
interference scattering and contains information about the local
packing structure. This scattering ring underwent dramatic changes
with the gradual addition, and subsequent removal, of liquid
solvent in the pores (see Fig.~\ref{fig:hiq}). No sharp
diffraction spots were observed upon rotation of the sample
through $90^\circ$ about the short axis of the pore, indicating
that the nanoparticle packing must be powder-like with only
short-range order.

    To interpret these particle-particle scattering results,
it was necessary to transform the scattering intensity from lab
(CCD detector) coordinates into the coordinates relative to the
nanopore axis as shown in Fig.~\ref{fig:transform}. The transform
is given by the constraints of the scattering geometry (see
Fig.~\ref{fig:geom}).  The wavevector transfer, $\vec{Q}=[Q_x,
Q_y, Q_z]$ is:
\begin{eqnarray}
\vec{Q}&=&\vec{k}_{inc}-\vec{k}_{scatt}\\
&=&\frac{k}{\sqrt{x^2+z^2+L^2}}\left[x,
L-\sqrt{x^2+z^2+L^2},z\right]\\
&\approx&\frac{k}{L}\left[x,0,z\right];\ \ L\gg r
 \end{eqnarray}
Transforming to coordinates of the nanopore (here we use
$\vec{q}$, lowercase for clarity), for $\theta_s\ll\pi/2$:
\begin{eqnarray}
q_z&=&\frac{\vec{q}\cdot\vec{m_z}}{{|m_z|}^2}\approx\frac{kz\cos{\theta_s}}{L}\\
q_y&=&\frac{\vec{q}\cdot\vec{m_y}}{{|m_y|}^2}\approx\frac{kz\sin{\theta_s}}{L}\\
q_x&=&\frac{\vec{q}\cdot\vec{m_x}}{{|m_x|}^2}\approx\frac{kx}{L}\\
q_r&\approx&\frac{k\sqrt{x^2+z^2{\sin}^2{\theta_s}}}{L};\ L\gg r;\
z\gg \Delta\cot{\theta_s}
\end{eqnarray}
Here, $\vec{x}=[x, y, z]$ ($\|\vec{x}\:\|=r$) and $\vec{m}=[m_x,
m_y, m_z]$ are the cartesian coordinates of the CCD detector (lab)
and nanopore, respectively, and $\Delta=\sqrt{x^2+z^2+L^2}$. The
wavevector transfer in the pore coordinates is denoted $\vec{q}$,
with magnitude $k=2\pi/\lambda$. In the above approximation, the
intersection of the cylindrical surfaces of constant $q_r$, and
the CCD plane, are ellipses. Fig.~\ref{fig:transform} shows
typical scattering intensity data (upper right) from $\Delta
T=30$~K (dry pores). The scattering intensity is transformed into
the $q_z, q_r$ coordinates of the nanopore (lower right). The
transform excludes data up to $\theta_s=10^\circ$ from the $q_z$
axis (aligned along the nanopore long axis) as a result of the
incident angle of the x-rays to the pore axis.  Images were taken
with the detector off-center to maximize the recorded $q$--range.
The triangular region at low q is a result of an attenuator
necessary to reduce the intense scattering of the powder
diffraction peaks associated with the hexagonal pore-pore packing.
The intense scattering concentrated along the $q_r$ axis was
mainly a result of scattering described by the pore form factor
which is narrow in $q_z$ due to the length of the pores.

    The intensity distribution along the nanoparticle scattering ring
changed significantly with $\Delta T$. For the dry pores, $\Delta
T=30$~K (Fig.~\ref{fig:hiq}, upper left) a sharp, strongly
asymmetric ring was present, brighter near the $q_z$ axis than the
$q_r$ axis.
 Here the data have been tiled to all four quadrants to simulate the
full ring for ease of viewing. The strong feature along the $q_r$
axis was due to the sum of the scattering of the individual pores.
The ring asymmetry indicates a structure that is preferentially
aligned along the nanopore axis. A model to explain this
scattering will be presented in the next section. As liquid was
added to the pores (adsorption curve) there was a general trend
that the ring became broader and its radius decreased (see
Fig.~\ref{fig:hiq}, $\Delta T=30~\mathrm{K}\rightarrow
8~\mathrm{K} \rightarrow 6~\mathrm{K}$). This was an indication
that the nanoparticle nearest neighbor spacing had increased and
the ordering was reduced in comparison with that of the dry case.
Upon gradual removal of the liquid (desorption curve) there was a
qualitatively different feature that appeared (see
Fig.~\ref{fig:hiq}, lower left, $\Delta T=12.4$~K).
For this value of $\Delta T$, the inner portion of the asymmetric
ring was nearly isotropic.  This isotropic scattering was an
indication of an additional structure of nanoparticle aggregates
or clusters.  With removal of nearly all of the liquid achieved at
the highest $\Delta T$, the scattering again showed a sharp, well
defined, asymmetric ring (almost identical to Fig.~\ref{fig:hiq},
upper left, $\Delta T=30$~K), indicating that the self-assembly
was reversible. The formation of the clusters indicated that there
was a form of hysteresis in the self assembly process as well as
in liquid adsorption and desorption.
\subsection{Nanoparticle Tiling Model} We propose in this section a simple
model to describe the scattering from monolayers of Au
nanoparticles along the walls of the cylindrical nanopores.
Motivation for a self-assembled cylindrical monolayer comes from
analogy with self-assembled nanoparticle monolayers on flat
substrates\cite{Narayanan04} and the attractive van der Waals
(vdW) forces between the particles and between the particles and
the walls of the nanopore.  In the Derjaguin approximation, where
the separation, $D$, is much less than the nanoparticle Au-core
radius, $R_s$, the vdW forces between neighboring nanoparticles,
$F_{np-np}$, and between a nanoparticle and the alumina pore wall,
$F_{np-wall}$, are:\cite{Israel92,Ohara95}.
\begin{eqnarray}
F_{np-np}&=&\frac{A_{nn}R_s}{12D^2}\\
F_{np-wall}&=&\frac{A_{nw}R_s}{6D^2}.
\end{eqnarray}
Here, $A_{nn}$ and $A_{nw}$ are the Hamaker constants for the
nanoparticle-nanoparticle interaction and nanoparticle-nanopore
wall respectively.  The Hamaker constants are reduced in the
presence of a mediating solvent, but the forces remain attractive.
Steric repulsions, due to the OT coating, prevent irreversible
nanoparticle aggregation.  The attractive vdW forces should
partially confine the particles to remain in a monolayer near the
walls of the pores.  Additionally, inter-particle ordering should
be reduced by pore wall roughness and nanoparticle
polydispersity.~\cite{Henderson96} The lack of strong diffraction
peaks confirmed that the nanoparticle order was short-range only.
There were two distinct length scales involved, the nanoparticle
diameter and the nanopore diameter, that differ by an order of
magnitude ($\sim3$~nm to $\sim30$~nm). Thus, there were two
independent scattering regimes: one for low angles,
$qR_c\stackrel{<}{\sim}1$, and another for higher angles,
$qR_s\stackrel{>}{\sim}1$, where $R_s$ is the average nanoparticle
core radius and $R_c$ is the average nanopore radius. For the low
angle scattering, which has already been discussed above in the
context of added liquid volume, the scattering due to interference
from neighboring nanoparticles was ignored for this reason. For
the scattering at higher angles, the structure factor associated
with the nanopore packing should have decayed to approximately
unity (since any pore-pore ordering must be short range in 2D).
This high $q$ regime may then be written as the sum of scattering
from the pores (as in the low $q$ region, but where now the
structure factor is unity) plus a term that describes the local
particle-particle scattering. The scattered intensity in the two
regimes is given by:
\begin{eqnarray}
I(q)&\approx&{|S_p(q)(F_s(q)+F_p(q))|}^2;\quad qR_c\stackrel{<}{\sim}1\\
I(q)&\approx&{|F_p(q)+S_{np}(q)F_{np}(q)|}^2;\quad
qR_s\stackrel{>}{\sim}1 \label{eq14}
\end{eqnarray}
Here, $S_p$ and $S_{np}$ are the structure factors associated with
the local packing of the nanopores and the nanoparticles
respectively. $F_p$, $F_s$, and $F_{np}$ are the form factors
associated with the scattering from the nanopores, an average
nanoparticle monolayer, and individual nanoparticles respectively.
Since the nanopores are about four orders of magnitude longer than
their diameter, the scattering associated with them will occur at
small $q_z$.  With the pores almost perpendicular to the direct
beam, the intersection of this scattering with the image plane
will be such that scattering associated with the pores will be
only along the $q_r$ axis.  The scattering associated with the
form factor of the pores is complicated by polydispersity,
roughness and the presence of liquid, thus it is treated only as
background scattering for this analysis.  Strong scattering away
from the $q_r$ axis must then be attributed to
nanoparticle-nanoparticle scattering.  Thus away from the $q_r$
axis, we may treat the 2nd term in equation (\ref{eq14}) as
dominant and the first term (and cross term) as a background.

    The nanoparticle form factor is described by:
\begin{eqnarray}
{|{F_{np}}|}^2&\propto&G(q,R_s,\Delta_R).
\end{eqnarray}
where $G(q,R_s,\Delta_R)$ is given by the Shulz
distribution\cite{Aragon76} for polydisperse spheres with radius
$R_s$ and $\Delta_R$ defined as the root mean square deviation.
For $qR_s\stackrel{<}{\sim}1$, the nanoparticle form factor is
given by the Guinier Law \cite{Glatter82}.
\begin{eqnarray}
G(q)\approx \exp(-q^2{R_s}^2/5)
\end{eqnarray}

\begin{figure}[tbp]
  \includegraphics[height=0.5\columnwidth]{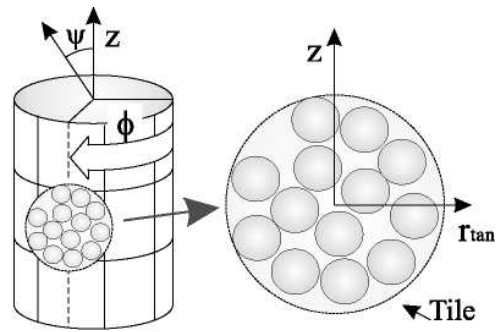}
  \caption{(Left) Schematic of the tiling of the cylindrical nanoparticle
  monolayer. The monolayer is broken up into small
  regions, or tiles, which are approximately flat.  The scattering is then
  powder averaged over all tiles at a given orientation $\phi$ from all pores.
  The total scattering is a sum over all such orientations, $\phi\in[0,2\pi]$.
  The azimuthal angle is  $\psi$ (not to be confused with the sample rotation angle $\theta_s$).}
  \label{fig:tile}
\end{figure}

    A ``tiling'' model for calculating $S_{np}(q)$, valid for
$qR_s\stackrel{>}{\sim}1$ is explored below. To model the
nanoparticle structure factor for high $q$ (where
nanoparticle-nanoparticle interference scattering dominates), the
monolayer of particles was broken up into small regions, or tiles,
that were approximately flat, as shown in Fig.~\ref{fig:tile}.
The approximation was made that the scattering between individual
tiles was independent and that the radius of curvature of the pore
is large compared with the inter-particle distance. The first
approximation is valid when the dimensions of the tiles are on the
order of the average correlation length (order parameter), $\xi$,
within the tile or smaller.

    The scattering was then powder--averaged over all tiles of a
given orientation $\phi$.  Thus the scattering intensity from a
particular set of tiles at a given $\phi$ was approximately the
same as the powder--averaged scattering from a flat monolayer of
nanoparticles with an order parameter $\xi$, and nanoparticle
nearest neighbor separation $d_{nn}$.  A (mathematically) simple
Lorentzian model was used to approximate the structure factor,
$S^{\phi}$, for each of these orientations:
\begin{equation}
{|S^{\phi}_{np}(q,\phi,\psi)|}^2\approx
\end{equation}
\begin{eqnarray*}
&&\frac{I_0\xi}{{1+\xi}^{2}{\left(q\sqrt{(\cos(\psi)-\cos(\phi)\sin(\psi))}-2\pi/d_{nn}\right)}^2}.
\end{eqnarray*}
Here $I_0$ is an adjustable scale parameter with $I_0/d_{nn}$
being proportional to the number of scatters and $\psi$ is the
azimuthal angle relative to the long axis of the nanopore (not to
be confused with the sample rotation, $\theta_s$).  Note that if
$\xi>d_{nn}$ then $\xi$ may be thought of as the correlation
length within the tiles, along the walls of the pore. For $\xi\leq
d_{nn}$ then the particles are better described as a 2D dilute,
gas-like phase. This analysis only considered the first order peak
due to the $q$--range of this experiment, but higher orders would
be treated similarly. The total structure factor, $S_{np}$, can be
calculated by integrating over all $\phi$.
\begin{eqnarray}
{|S_{np}(q,\theta)|}^2&=&\int^{2\pi}_0{|S_{np}^{\phi}(q,\phi,\theta)|}^2d\phi
\end{eqnarray}
    The above structure factor, along with the nanoparticle form
factor given by the Shulz distribution, was used to fit the high
$q$ data of the Au-core nanoparticles in the nanoporous alumina.
The CCD data were first flat-field corrected, then background
(taken with the sample rotated out of the beam) was subtracted to
account for scattering from our environmental chamber windows plus
scattering from the toluene vapor. The data were then transformed
into the $q_r, q_z$ coordinates of the nanopore as described
above.  All of the data sets (at different $\Delta T$) fit well
with the above described tile model with a monotonic decaying
background (which included scattering from both the porous
membrane and the liquid) with the exception of the data at $\Delta
T=12.4$~K (desorption curve) where an additional isotropic
scattering term was necessary for good agreement.  This
cross-section of the isotropic scattering ring was fit using a
Lorentzian with similar parameters as for the monolayer ($\xi,
I_0,$ and $d_{nn}$).  For this model, the number of scatters
should be proportional to $I_0/d_{nn}^2$.

\begin{figure}[tbp]
\includegraphics[width=1\columnwidth]{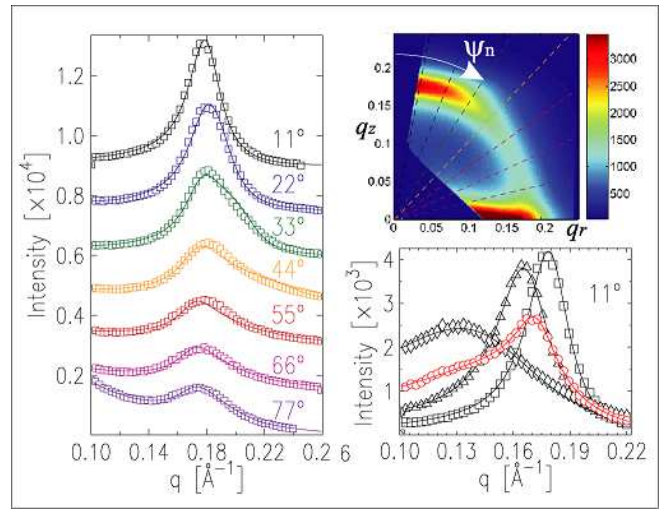}
\caption{(Color)(Right Top) Schematic of representative radial
slices (dashed lines) taken for fitting with the tile model with
$\psi_n=11\times n^\circ$, $n$ an integer from 1 to 7.  (Left)
Data ($\Box$) at $\Delta T=30$~K from each slice (color coded with
dashed lines) along with fits (solid lines) from the tile model
plus independent monotonically decaying, positive backgrounds.
Error bars are smaller than data symbols.  (Bottom Right) Data
(all at $\psi_n=11^\circ$) from four different $\Delta T$:
$\Box=30$~K, $\bigtriangleup=8$~K(absorption),
$\lozenge=4.4$~K(saturated), and $\bigcirc=12.4$~K(desorption);
solid lines are fits.  $\Delta T=12.4$~K is qualitatively
different and requires an isotropic component in addition to the
tile model to fit properly.}\label{fig:shellmulti}
\end{figure}

    For each data set of a given $\Delta T$,
a series of representative radial slices was chosen at
$\psi_n=11\times n^\circ$ from the $q_z$ axis ($n$ an integer from
one to seven), and intensity as a function of $q$ was graphed for
each slice (see Fig.~\ref{fig:shellmulti}).
Each radial slice was fitted independently with the tiling model
plus a monotonically decaying background. As stated above, the
main contribution to the background was from the nanopore membrane
which increased as $\psi_n$ approached 90$^\circ$. This was seen
in the fits and thus a slice at 88$^\circ$ was not used due to the
fact that the background was greater than the particle-particle
scattering. After fitting all of the slices independently, average
parameters and uncertainties for each $\Delta T$ were calculated
from the mean and standard deviation of each of the three fit
parameters:  $I_0$, $\xi$, and $d_{nn}$.  Since the uncertainties
were small in comparison to the average value of the parameter,
the average parameters were treated as "best fit" or global
parameters for each $\Delta T$.

\begin{figure}[tbp]
  \includegraphics[width=1.0\columnwidth]{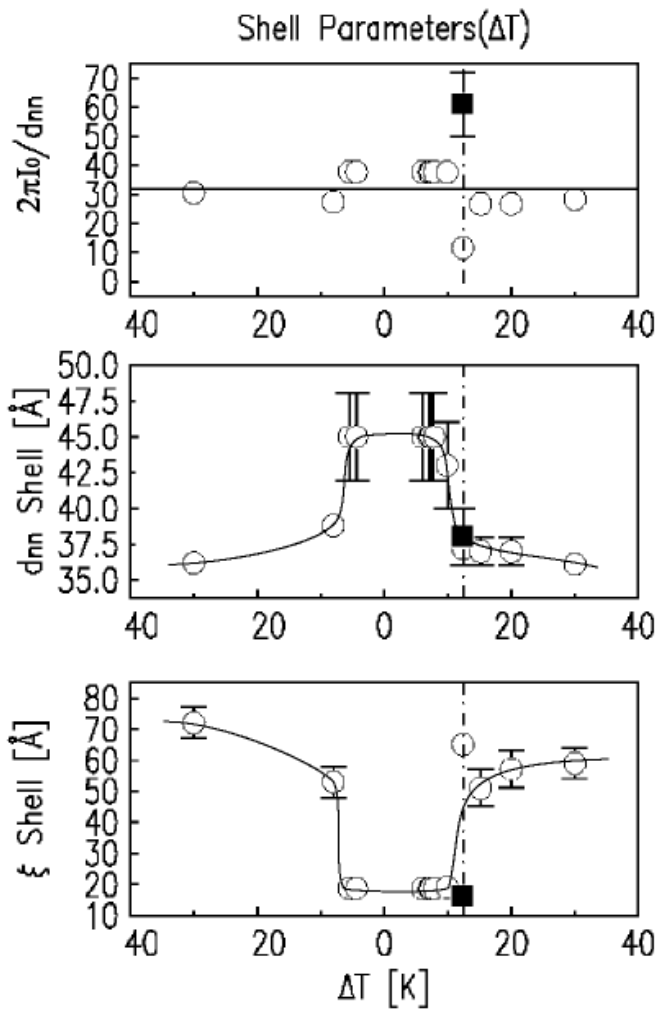}
  \caption{Plots of the "best fit" tile model ($\bigcirc$) and isotropic ($\blacksquare$)
  parameters and uncertainties as a function of $\Delta T$ (data to the left of zero are from adsorption curves,
  to the right are from desorption curves).  The physical parameters are (top to bottom):
  number of scatters $\propto I_0/d_{nn}$, nearest neighbor spacing $d_{nn}$, and order parameter $\xi$.
  Lines are added as a guide to the eye.}\label{fig:param}
\end{figure}

    The three "best fit" parameters from the tiling model as a function
of $\Delta T$ are plotted with uncertainties in
Fig.~\ref{fig:param}. Also plotted in each of the three plots are
the parameters from the isotropic component at $\Delta T=12.4$~K
($\blacksquare$).  The $\Delta T$ axis is read from left to right;
to the left of zero are adsorption (cooling) data, to the right of
zero is desorption (heating) data.

The data indicated that the number of nanoparticles in the
monolayer, probed by the scattering volume, remained constant as
liquid was added and subsequently removed (see
Fig.~\ref{fig:param}, top) with the exception of $\Delta T=12.4$~K
where an additional isotropic component was necessary. For this
data point, the number of particles in the monolayer decreased. It
is likely that some of the missing particles went into the
formation of the isotropic clusters.  The fact that this data
point is higher than the average monolayer number might be
indicative that the clusters fill the volume more effectively than
the particles in the monolayer.

  The formation of these clusters is a surprising phenomenon.
Additional isotropic scattering was seen only during desorption
and then disappeared upon removal of additional liquid. The
formation of clusters, or aggregates may have been linked to the
progression of a spherical meniscus through the pores, starting
from the ends.\cite{Cohan38}  As the meniscus travelled through
the pore, it would tend to drag particles with it. Another
possibility would be the formation of bubbles within the liquid.
Upon complete removal of liquid from the system these clusters
would dissolve and reassemble into a monolayer structure on the
nanopore wall. This is indicated in the data by the disappearance
of the isotropic scattering as $\Delta T$ is increased.

    The plots in  Fig.~\ref{fig:param}, middle and bottom,
demonstrate that with the addition of liquid there was a shift to
larger nearest neighbor separation distance with a maximum of
4.5~nm (compared to 3.6~nm for dry), accompanied by a decrease in
the order (in the plane of the cylindrical monolayer).  Upon
removal of the liquid solvent, the particle separation decreased
again to the dry value, and the ordering increased back to almost
the initial dry value. For relative added liquid solvent volume
amounts up to about 0.5 (normalized to saturation; see both
Fig.~\ref{fig:lowq} and Fig.~\ref{fig:param}) the order parameter
was about twice that of the nearest neighbor spacing and may be
interpreted as an in-plane correlation length. For higher relative
amounts of solvent, $V_{liq}>$0.5, the order parameter was less
than the nearest neighbor spacing, indicating a 2D dilute gas-like
phase.

    The value of 3.6~nm for the nearest neighbor spacing for the
dry system (large $\Delta T=30$~K) was less than
$2(R_s+t_{OT})\approx 4.8$~nm, where $t_{OT}\approx 1.2$~nm is the
thickness of the OT shell, indicating interdigitation of the shell
ligands of neighboring particles.  As liquid was absorbed into the
organic shells (with decreasing $\Delta T$) the vdW attraction
between nanoparticles would decrease\cite{Ge00, Ohara95} relative
to the dry system, while the osmotic pressure\cite{Saunders04}
between the ligand chains increased, driving the particles apart.
This separation increase may have been further facilitated by gaps
or voids in monolayer coverage. As liquid filled these gaps, the
nanoparticles (undergoing thermal motion) could move into this new
volume. Removal of the liquid by increasing $\Delta T$ reduced the
repulsive osmotic pressure between the particles and increased the
attractive vdW forces, reducing the particle-particle separation.

\section{CONCLUSIONS}
Low angle measurements established the relative amount of liquid
in the nanopores as a function of $\Delta T$.  A capillary filling
transition occurred between 8~K and 12~K, which was about four
times less saturation than what is expected via the Kelvin
equation for toluene absorption in pores with no nanoparticles. In
addition to the shift, marked hysteresis was also observed.

Fits to the data indicated that the nanoparticles assembled in a
cylindrical monolayer aligned along the pore. From the geometry of
this structure we conclude that the particles were near the walls,
which was physically sensible due to the vdW attraction of the
nanoparticles to the nanopore wall. As liquid was added by
reducing the relative chemical potential $\Delta\mu$, most of the
particles remained in the monolayer structure.  Also, the nearest
neighbor separation distance increased and the correlation length
within the monolayer decreased.  The process was reversible: upon
removal of the liquid, the nanoparticle nearest neighbor distance
decreased to the initial dry value and the ordering increased to
almost the dry value.

    In addition to the cylindrical monolayer,  there was evidence
of the formation of isotropic clusters during desorption of the
liquid solvent.  This phenomenon could have been related to the
process by which the liquid emptied from the pore, namely that it
was likely to occur from the ends of the pores.  This might also
explain why this structure was not seen during the adsorption
cycle.\\

 We thank Richard Schalek for help with preparing ultra-microtome
TEM samples.  We also thank Professor Milton Cole for many helpful
discussions. This work was supported by the National Science
Foundation Grant No. 03-03916.  ChemMatCARS Sector 15 is
principally supported by the National Science
Foundation/Department of Energy under grant number CHE0087817. The
Advanced Photon Source is supported by the U.S. Department of
Energy, Basic Energy Sciences, Office of Science, under Contract
No. W-31-109-Eng-38. Use of the Advanced Photon Source was
supported by the U. S. Department of Energy, Office of Science,
Office of Basic Energy Sciences, under Contract No.
W-31-109-Eng-38.


\bibliographystyle{unsrt}

\begin{thebibliography}{99}
\bibitem{volume}
This is a result of the approximation of the form factor for small
wavevector transfer: $|F(q)|^2=|\int{d\vec{x}
\rho(\vec{x})\exp(i\vec{x}\cdot\vec{k})}|^2\approx |\int{d\vec{x}
\rho(\vec{x})}|^2 = V^2.$
\end{thebibliography}

\end{document}